\documentclass[9pt, conference]{IEEEtran}
\IEEEoverridecommandlockouts

\usepackage{graphicx} 
\usepackage{float}
\usepackage{placeins}
\usepackage{algorithm} 
\usepackage{algpseudocode} 
\usepackage{subcaption} 
\usepackage{xcolor} 
\usepackage{pgfplots}
\usepackage{hyperref}
\usepackage{cite}
\usepackage{amsmath,amsfonts,amssymb}
\usepackage{enumitem}

\setlist[itemize]{align=parleft,left=0pt..1em}

\newcommand*\circled[1]{\tikz[baseline=(char.base)]{
            \node[shape=circle,draw,inner sep=1pt] (char) {#1};}}

\usepackage{amsthm}
\usepackage[utf8]{inputenc}
\usepackage[english]{babel}
\usepackage[T1]{fontenc}
\theoremstyle{definition}
\newtheorem{observation}{Observation}

\newcommand{\blockcomment}[1]{}

\IEEEpubid{\makebox[\columnwidth]{ 979-8-3315-2710-5/25/\$31.00 $
\copyright$2025 IEEE \hfill{ \hspace{\columnsep} \makebox[\columnwidth]{ }}}}

\begin{document}
\pagestyle{plain}
\pagenumbering{arabic}

\title{ECLIP: \underline{E}nergy-efficient and Practical \underline{C}o-\underline{L}ocation of ML \underline{I}nference on Spatially \underline{P}artitioned GPUs}
\author{\IEEEauthorblockN{Ryan Quach\IEEEauthorrefmark{1}, Yidi Wang\IEEEauthorrefmark{2}, Ali Jahanshahi\IEEEauthorrefmark{1}, Daniel Wong\IEEEauthorrefmark{1}, Hyoseung Kim\IEEEauthorrefmark{1}}
\IEEEauthorblockA{
\IEEEauthorrefmark{1}University of California, Riverside\\
\{rquac004, ajaha004, danwong, hyoseung\}@ucr.edu}
\IEEEauthorblockA{\IEEEauthorrefmark{2}Santa Clara University \\
ywang49@scu.edu}

}

\date{2025}
\maketitle
\begin{abstract} 
As AI inference becomes mainstream, research has begun to focus on improving the energy consumption of inference servers. Inference kernels commonly underutilize a GPU's compute resources and waste power from idling components. To improve utilization and energy efficiency, multiple models can co-locate and share the GPU. However, typical GPU spatial partitioning techniques often experience significant overheads when reconfiguring spatial partitions, which can waste additional energy through repartitioning overheads or non-optimal partition configurations. In this paper, we present ECLIP, a framework to enable low-overhead energy-efficient kernel-wise resource partitioning between co-located inference kernels. ECLIP minimizes repartitioning overheads by pre-allocating pools of CU masked streams and assigns optimal CU assignments to groups of kernels through our resource allocation optimizer. 
Overall, ECLIP achieves an average of 13\% improvement to throughput and 25\% improvement to energy efficiency.

\end{abstract}

\section{Introduction} 
Inference servers have emerged as a critical component as ML/AI becomes increasingly prevalent in IoT applications from smartphones to self-driving vehicles. While these systems rely heavily on GPUs to execute inference computations of ML models, individual inference requests often do not utilize all GPU compute resources, e.g., Compute Units (CU) in an AMD GPU~\cite{scaleserve} and Streaming Multiprocessors (SM) in an Nvidia GPU~\cite{gpulet},~\cite{elasticroom}. Therefore, underutilized GPUs waste energy by having many idling CUs or SMs that cannot be power-gated due to architectural constraints~\cite{kandiah2021accelwattch,9622409,cofris,zamani2021icap,jahanshahi2020gpu}. 
This offers an opportunity for significant energy and performance optimizations by sharing a GPU among concurrent ML models. 

The effectiveness of an inference server is primarily assessed by its ability to meet Quality of Service (QoS), such as throughput and tail latency.
When multiple requests are scheduled concurrently on a GPU for throughput, tail latency may increase significantly due to interference of shared compute and memory resources. 
To improve QoS of co-locating models, spatial partitioning allocates distinct portions of GPU internal compute resources to kernels within a request, lessening contention. However, even the state-of-the-art inference servers employing spatial partitioning~\cite{classsearch, warperslicer, improvedscale, KRISP, elasticroom} still suffer from the following key limitations. 


Prior works trade-off the granularity of spatial partitions with the overheads of partition reconfiguration. Due to high overheads of Nvidia's MPS and MIG spatial partitioning, many prior works partition at the coarse granularity of entire models~\cite{gslice, pariselsa, gpulet}. However, this approach leaves significant GPU under-utilization since compute resource requirements vary kernel-by-kernel within an inference request. To address this, kernel-grain spatial partitioning was proposed with architectural extensions to avoid overhead of repartitioning at every kernel~\cite{KRISP}. However, it requires hardware modifications that are impractical in real GPUs, leading to significant energy and performance overheads. 



\textbf{Our Contribution.}
This paper introduces ECLIP, a framework for \textit{energy-efficient} and \textit{practical} co-location of ML inference on spatially partitioned GPUs. ECLIP introduces both \textit{scheduling} and \textit{resource allocation} approaches to achieve effective kernel-wise spatial partitioning with minimal spatial repartitioning overheads on real-world GPUs. 

We first characterize the overheads of CU masking and model co-location through carefully designed experiments on an AMD GPU. Based on these observations, we propose a lightweight runtime scheduler that achieves kernel-grain partitioning without substantial overhead, by creating pools of pre-allocated CU mask streams. To guide scheduling decisions, we formulate a resource allocation model that determines energy-cognizant CU allocation for \textit{groups of kernels} within an inference request while accounting for co-located ML model scenarios and their associated slowdowns. By limiting repartitioning events and making allocation decisions at kernel group granularity, we achieve the benefits of per-kernel repartitioning with minimal energy waste on real GPU systems.

Experimental results demonstrate that ECLIP can achieve up to 21\% increase in throughput and on average a 25\% improvement in energy efficiency, while maintaining tail latency in an acceptable range. 
ECLIP is integrated into the AMD ROCm runtime, remaining transparent to PyTorch and other ML inference backends, enabling energy and performance benefits without any model or PyTorch modifications.


\section{Background}
\label{sec:background}

\subsection{AMD GPU Hardware}
AMD GPUs consist of multiple Compute Units (CUs) \cite{amdhwbasics}, which are equivalent to Streaming Multiprocessors (SMs) in Nvidia terminology. Multiple CUs are organized into Shader Engines (SEs), typically in clusters of 15 CUs for the MI50 architecture. 
Unlike Nvidia's MPS or MIG, which applies spatial partitions at the process level, AMD's \textit{CU Masking} feature within the ROCm runtime enables spatial partitioning at a finer granularity of streams by specifying which CUs to execute kernels on. This provides more flexible resource allocations for inference servers. 

Additional architectural details on AMD GPUs can be found in technical whitepapers \cite{GCNarchitecture, Polarisarchitecture, CDNA2architecture} and academic papers \cite{Alternatives}. 
AMD's ROCm documentation \cite{ROCmDoc} and the HSA Foundation's Programmer Manual \cite{HSADoc} provide essential information about AMD's runtime backend. 



\subsection{GPU Energy-efficiency Techniques}\label{sec:gpu_energy}
GPUs can save power through DVFS and resource scaling techniques. 
While DVFS can scale frequency and voltage to save dynamic power, its impact is limited by underutilization of idle resources which typically suffer from leakage power.
Resource scaling, such as limiting workload to a subset of CUs, can in theory save power by allowing unused resources to be power-gated to save leakage power. However, power gating techniques in production GPUs are limited to clock-gating or gating of the entire GPU~\cite{kandiah2021accelwattch,towardsenergy,cofris,abdel2013warped,jahanshahi2020gpu}. 
Hence, CU masking alone does not lead to significant energy-efficiency gains. To address under-utilization, it is favorable to co-locate and increase the utilization of the GPU to improve energy efficiency. 




\subsection{Limitations of GPU Partitioning for ML Inference}


ML inference tends to underutilize GPUs~\cite{gpulet,KRISP,gslice,elasticroom,pariselsa,WattWiser,cofris,xiang2019pipelined}. To improve utilization, spatial co-location of ML models allow concurrent ML inference requests to compute on different GPU resources. However, there are two challenges. First, ML models must be right-sized in order to use the minimal amount of resources while maintaining performance. Second, GPU spatial partitioning mechanisms can incur high overhead.

\subsubsection{Model-grain Right-Sizing}

ML models are typically sized to use minimal resources while meeting a certain quality-of-service target, typically 3x the tail latency when running in isolation~\cite{pariselsa,gpulet}. The amount of resources dedicated to an ML model is generally identified at the ``knee'' of the resource vs. latency curve.
Since GPU spatial partitioning incurs high overhead, many prior works perform right-sizing ML models at the granularity of the entire model. 
However, individual kernels within an inference request require different amounts of resources, resulting in idling CUs that waste energy. 
Prior works like PARIS$+$ELSA~\cite{pariselsa}, GPUlet~\cite{gpulet}, and GSlice~\cite{gslice} all rely on model-grain right-sizing for GPU partitions through MPS or MIG on Nvidia GPUs. These approaches create shadow partitions in the background and hot-swap the ML inference once it completes loading.

\subsubsection{Kernel-grain Right-Sizing}
At finer granularities, ML models can be right-sized at the kernel level~\cite{elasticroom,KRISP}, requiring per-kernel spatial repartitioning. However, with ML inference kernels often lasting as little as microseconds, the spatial repartitioning overhead can exceed the kernel's actual runtime. Furthermore, while AMD GPUs contain a mechanism for CU masking, applying CU masks every kernel requires code modifications to Pytorch.
To get around this, ElasticRoom~\cite{elasticroom} employs compiler code-generation of an array of GEMM kernels with varying resource requirements. Then at runtime, a scheduler selects appropriately sized kernels to run. However, this approach is limited to only GEMM kernels and requires significant runtime modifications that are not compatible with  ML frameworks like PyTorch. 
KRISP~\cite{KRISP} proposes microarchitectural extensions to support kernel-level CU masking with low-overheads. This enables kernel-wise right-sizing of kernels without any code modification of ML frameworks. However, such mechanisms cannot be used for existing GPU hardware due to the lack of microarchitectural extensions, as seen from their evaluation in an emulation environment.  

A recent work proposed libsmctrl~\cite{HWPart}, a library that enables SM Masking on Nvidia GPUs using undocumented CUDA debugging callback API to modify hardware resource mask associated with each kernel. libsmctrl is able to enforce a SM Mask globally, for specific streams, or for individual kernels. This would incur a libsmctrl API call before every kernel launch and would require modifications to existing ML frameworks to support this, which again is not very practical. 



\subsection{Challenges towards practical kernel-level spatial partitioning of ML inference}

Existing GPU spatial partitioning support for ML inference servers are currently limited by the overheads of spatial partitioning or by the impracticalities of program changes to ML frameworks and custom microarchitectural changes. 
Therefore, a key to providing \textit{practical} spatial partitioning for co-located ML inference servers is to carefully manage the overheads of spatial partitioning. In order to do so, we need to deeply understand the intricacies between workload latency, co-location and hardware CU masking overheads.  


In this work, we build off AMD CU masking mechanism to provide a practical spatially partitioned ML inference server, called ECLIP. Our insight is that while partitioning on a per-kernel basis is extremely costly, we aim to limit the amount of spatial partitioning overhead by partitioning at \textit{groups of kernels}. To achieve this, we characterize CU masking properties in AMD GPUs to build analytical models (for inference latency and overheads) and resource allocation optimization to achieve practical kernel-wise co-location of ML inference on real GPUs. 

\section{Analyzing Hardware Characteristics}\label{sec:hardware_characteristics}
Our analysis is based on an AMD Radeon Instinct MI50 GPU, containing 60 CUs, organized as 4 SEs of 15 CUs each. We investigate AMD CU masking due to enforcing partitions at the granularity of streams and unlocks repartitioning \textit{within} model inference.
Our experiments aim to investigate overheads of resource partitioning, which can lead to subpar performance and energy efficiency. 

\begin{figure}
\begin{subfigure}{.49\linewidth}
    \centering
    \includegraphics[width=1\linewidth]{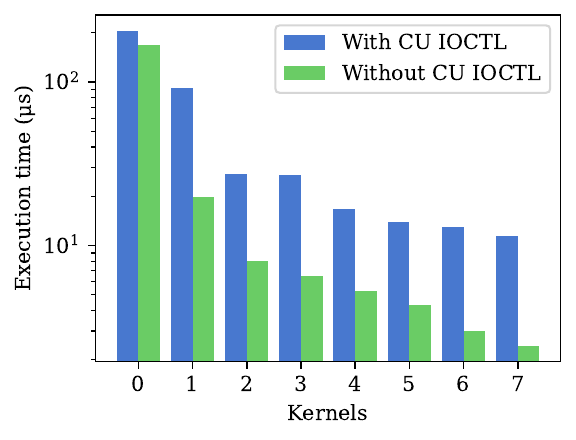}
    \caption{Latency of 8 longest \texttt{albert} kernels.}
    \label{fig:cucost}
\end{subfigure}
\begin{subfigure}{.49\linewidth}
    \centering
    \includegraphics[width=1\linewidth]{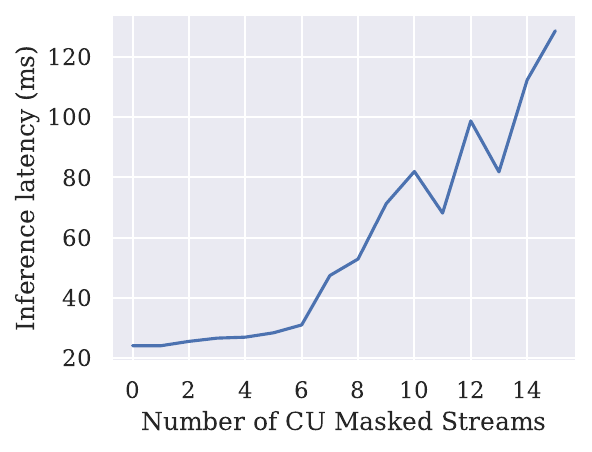}
    \caption{Number of CU-masked Streams vs. Inference Latency.}
    \label{fig:streamvslag}
\end{subfigure}
\caption{CU masking timing characteristics.}
\end{figure}

The first experiment is to examine the CU mask API overhead. This is done by instrumenting ROCm~\cite{ROCmDoc}, which is AMD's runtime stack equivalent to Nvidia's CUDA, binding the user space kernel code with the device driver.  
Within ROCm, a runtime API sets the hardware queue’s CU mask through an IOCTL system call. Specifically, this IOCTL call accounts for the majority of CU masking repartitioning overhead and leads to wasteful GPU idling time. To examine this overhead, we update the CU mask of every single kernel within an inference pass for the \texttt{albert} model~\cite{albert}.

\begin{observation}
CU masking IOCTL calls have high and unpredictable timing overheads.
\end{observation} 

We observed a significant slowdown from calling the CU mask IOCTL call every kernel.\label{obs:ioctl} 
In Figure \ref{fig:cucost}, we show the 8 longest running kernels averaged over 100 inference requests, and compare their latency with or without per-kernel CU masking. 
In the worst scenario, the CU IOCTL overhead is 55.4~\textmu s, over 2x the kernel execution time of 19.8~\textmu s.

    

The next experiment explores whether the number of CU masked streams introduce additional overheads. This is important since model co-location requires multiple concurrent streams, each with unique CU masks. We create an increasing number of CU masked streams (each with 3 unique CUs) and distribute inference kernels across these streams in a round-robin fashion. HSA barrier packets in ROCm are used to ensure that kernels execute in correct sequence to maintain their data dependencies (see Sec.~\ref{sec:runtime_scheduler} for details).




\begin{observation}
The GPU can accommodate a finite number of CU masked streams before experiencing slowdown due to queue oversubscription.\label{obs:hsa_queues}
\end{observation}

As shown in Figure~\ref{fig:streamvslag}, with up to 7 CU masked streams, there is relatively stable performance. However, latency significantly and unpredictably rises with each subsequent CU masked stream.

This likely stems from the GPU's limited number of hardware queues. Whenever a CU masked stream is created, there is a corresponding command queue created within the software runtime, i.e., HSA queues in ROCm. While these queues help the concurrency of multiple kernels, 
performance issues arise once the number of HSA queues exceeds the number of physical hardware queues. In such cases, the GPU must \emph{oversubscribe} HSA queues to hardware queues, where the hardware periodically switches between all allocated HSA queues~\cite{Oversub}, resulting in longer latencies. Similar performance issues have been reported on the latest generations of GPUs, including AMD RX9700 XT. 

The above experiment informs our design decision. Including a default stream in ROCm, the total number of streams we can support on our GPU is 8, matching the number of hardware queues~\cite{runlistoversub,oversubdoc}. 
Therefore, to prevent significant oversubscription overheads, we need to limit our creation of CU masked streams to 7.

\pgfplotstableread[row sep=\\,col sep=&]{
ckerns2 & time \\
1 & 262.391 \\
2 & 260.538 \\
3 & 261.557 \\
4 & 260.558 \\
5 & 336.458 \\
6 & 385.754 \\
7 & 421.492 \\ 
8 & 447.246 \\
9 & 520.802 \\
10 & 578.731 \\
11 & 626.670 \\
12 & 664.911 \\
}\mydata

\FloatBarrier

\section{ECLIP Framework}\label{sec:CLIP_framework}
\begin{figure}[!t]
    \centering
    \includegraphics[width=1\linewidth]{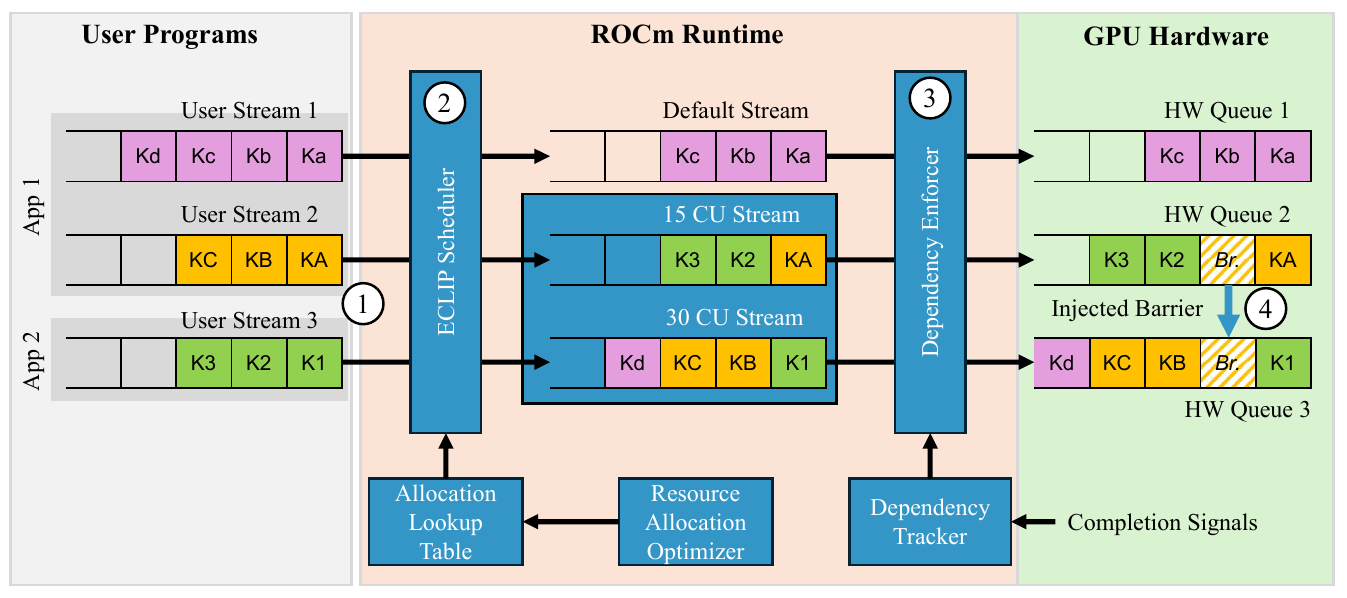}
    \caption{ECLIP Overview. Additions to ROCm in blue.}
    \label{fig:overview}
\end{figure}

Based on the observations made in Section~\ref{sec:hardware_characteristics}, we propose ECLIP, a framework for \textit{practical} and \textit{energy-efficient} co-location of inference kernels. ECLIP aims to trade off spatial repartitioning granularity with repartitioning overheads. 
ECLIP achieves effective kernel-granular spatial partitioning by (i) creating pools of streams with pre-defined CU masks, (ii) a runtime scheduler that intercepts and redirects kernels to appropriate CU-masked streams while maintaining inter-kernel data dependencies, and (iii) a resource allocation optimizer that assigns optimal CU assignments for \textit{groups of kernels} while considering co-location scenarios, and other hardware constraints. 
Figure~\ref{fig:overview} shows an overview of ECLIP, with require modifications to the ROCm runtime in blue boxes.



\subsection{Runtime Scheduler}\label{sec:runtime_scheduler}


\noindent\textbf{Pre-allocated CU Masked Streams:} 
ECLIP pre-allocates \emph{seven} CU-masked streams in the HIP (Heterogeneous-Compute Interface
for Portability) backend, ROCm’s user-facing layers. This approach avoids costly CU masking IOCTL calls at runtime, as previously shown in Obs. 1. The number of CU-masked streams is determined based on Obs. 2, which shows that creating more than 7 streams would introduce significant performance degradation.

ECLIP follows the common ML inference server architecture where each distinct ML model is processed by a designated worker thread that handles all inference requests for its model.\footnote{In the rest of the paper, we use models and workers interchangeably.}
For these workers, streams are created with CU increments of 15, as each Shader Engine (SE) on the GPU has exactly 15 CUs. This ensures that CUs are assigned in complete SE units, thereby minimizing interference between co-located workers. This also prevents performance loss when CUs are unevenly distributed across SEs~\cite{WorstPractices}. For example, if we have 2 workers, each worker will have their own set of CU masked streams with 15, 30, and 45 CUs. Worker 1's CUs are allocated to the following SEs: 15 CU (SE 1), 30 CU (SE 1, 3), 45 CU (SE 1, 2, 3).  Worker 2 will be assigned with: 15 CU (SE 4), 30 CU (SE 2, 4), 45 CU (SE 2, 3, 4). In both workers, the 60 CU allocation is the default stream, which keeps us below the hardware queue limit. For 3 co-located workers, each worker has their own set of 30 and 45 CU masked streams, with a default shared stream of 60 CUs.

\smallskip
\noindent\textbf{Redirecting Kernel Dispatch to Pre-allocated CU Masked Streams}: As shown in Figure \ref{fig:overview}, ECLIP modifies the HIP backend's kernel dispatch process (\circled{2}). It intercepts kernels and uses a lookup table to determine which CU-masked stream each kernel should use. The incoming kernel (\circled{1}) is then removed from its original stream and redirected to the selected CU-masked stream. This lookup table is generated by our resource allocation optimizer (Sec.~\ref{sec:optimizer}), which aims to maintain exclusive streams with minimal sharing.

\smallskip
\noindent\textbf{Enforcing Data Dependencies: } 
By default, data dependency between kernels is enforced within a user stream since kernels are executed one-by-one in FIFO order. However, when ECLIP redirects kernels to different pre-allocated CU masked streams, we need an additional mechanism to enforce the data dependencies between kernels in different pre-allocated CU masked streams. 
To address this issue, ECLIP introduces a dependency enforcement mechanism in the ROCclr library's dispatch process (Figure~\ref{fig:overview} \circled{3}). 
Hence, we dynamically inject HSA barrier packets to enforce dependencies between kernels that originate from the same user stream but are assigned to different CU-masked streams (and therefore, dispatched to different hardware queues).
A barrier packet is required when (i) the kernel has a dependency on a previous kernel from the same stream, and (ii) that previous kernel has not yet completed execution. 

To implement this, our dependency enforcer maintains a two-dimensional vector tracking kernel execution status: one dimension for user streams and the other for kernel completion signals.\footnote{A completion signal is the runtime's way to determine whether a kernel has finished. If it is 0, it is still running. If it is 1, the old kernel was completed. We also use this same signal for profiling.} For each incoming kernel, the enforcer checks the completion signal of the previous kernel from the same stream. If that kernel is still running, the enforcer creates and dispatches a barrier packet dependent on its completion signal, followed by the new kernel (\circled{4}). Otherwise, the new kernel is dispatched immediately. In both cases, the new kernel's completion signal is logged for future reference.

Although barriers are necessary for dependency management, they can introduce significant overhead as the command processor must track dependency signals for each barrier and only release blocked kernels once these signals are satisfied.
The excessive use of barriers can also lead to unpredictable slowdown. In our tests, the overhead could exceed the actual kernel execution time. We address this challenge through barrier packet budgeting in our resource allocation modeling and optimization.  

\subsection{Resource Allocation Optimizer}
\label{sec:optimizer}
To further minimize the impact of fine-grain spatial partitioning, ECLIP assigns resources to \textit{groups of kernels}. ECLIP introduces a resource allocation optimizer that determines \textit{how many} CUs a group of kernels will be allocated, and \textit{when} to switch CU partitions (which determine how kernels are grouped).

\begin{figure}[t]
    \centering
    \includegraphics[width=1\linewidth]{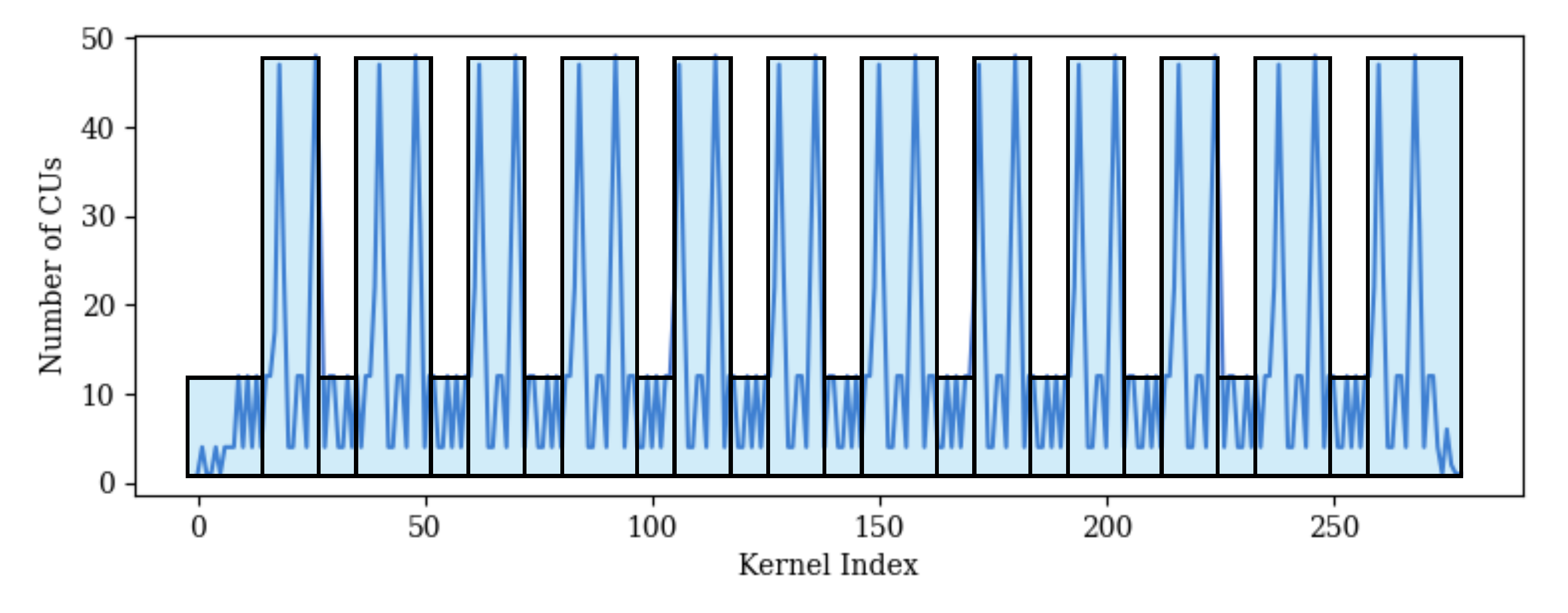}
    \caption{CU allocations for \texttt{albert} by ECLIP's optimizer}
    \label{fig:modeledband}
\end{figure}

\smallskip
\noindent\textbf{Minimum CU Threshold Profiling:}
We profile each kernel's \emph{independent} execution time with a single request on the GPU and identify a ``threshold'' per kernel that is the minimum number of CUs needed without experiencing noticeable slowdown. As shown in Figure~\ref{fig:modeledband} (light blue line), different kernels within a request can have widely varying minimum CU thresholds, creating opportunities for efficient resource sharing.

\smallskip
\noindent\textbf{CU Switching Overhead:}
Frequent switches between CU configurations introduce significant overhead, as explained in Sec.~\ref{sec:runtime_scheduler}. Each switch requires a barrier packet to maintain data dependencies, but excessive barrier packets can significantly degrade performance. Therefore, our optimizer must balance the benefits of fine-grained CU allocation against the overhead of configuration switches by considering \textit{groups of kernels} together rather than optimizing each kernel's CU configuration independently. Figure~\ref{fig:modeledband} (blue boxes) shows an illustrative example of how our optimizer performs CU allocation to groups of kernels.



\smallskip
\noindent\textbf{Fairness among Workers for Energy Efficiency:}
In co-located scenarios where multiple workers handle inference requests concurrently, our optimizer must ensure fairness while maximizing overall GPU utilization. This fairness is critical because disparate completion times among workers lead to idling CUs, wasting energy due to the lack of CU-level power gating as explained in Sec.~\ref{sec:gpu_energy}. To address this issue, we associate one identically weighted objective function per worker, where each objective function aims to minimize the execution time of that worker's kernels. 

\smallskip\noindent\textbf{Optimization Formulation:}
Our optimization goal is to minimize execution time across all workers while maintaining fairness and minimizing slowdowns. For each kernel, the model must select a CU configuration from a predefined set of configurations that align with the SE organization (increment of 15 CUs. Kernel execution time is influenced by three factors: (i) the number of CUs allocated to it, (ii) the slowdown from co-location with other workers' kernels, and (iii) the overhead of CU configuration switches.

Factor (i) is directly determined from profiling data. Factor (ii) is modeled based on CU sharing because the slowdown experienced by workers tends to be proportional to their CU overlap, e.g., higher slowdown with more shared CUs. Hence, we compute this by tracking the average number of CUs used by each worker and determining potential overlap between workers. To minimize factor (iii), we introduce \emph{barrier packet budget} that constrains the total number of CU configuration switches allowed within each worker's request.

The mathematical formulation of our model is as follows. 
The model's objective function is to minimize the summation of kernel execution time for all kernels in each worker, with equal weights for all workers for fairness:
\begin{center}
$\forall w, \text{minimize} \sum_{k \in w} e_k$
\end{center} 
where $w$ is a worker and $e_k$ is the estimated execution time of a kernel $k$ in co-located scenarios. 

Let $x_{k,c}$ be the main decision variable representing whether the CU configuration $c \in \text{configs}$ is selected for the kernel $k$, where $\text{configs}=\{15, 30, 45, 60\}$ is the set of available CU configurations. $x_{k,c}=1$ if $k$ uses the configuration $c$, and $x_{k,c}=0$ otherwise. With $x_{k,c}$, we introduce the following constraints for configuration selection: 
\begin{itemize}
    \item $\forall k\in w, \sum_{c\in \text{configs}} x_{k,c} = 1$: This forces the model to assign only one configuration to each kernel.
    \item $\text{switchTotal}_w = \sum_{k\in w}\sum_{c \in \text{configs}} |x_{k,c}- x_{k-1,c}|/2$: This counts the total number of switches occured for a worker $w$. The term $|x_{k,c}- x_{k-1,c}|$ detects a switch between consecutive kernels as their $x$ values differ when using different configurations. Division by 2 is to correct for double counting.
    \item $\text{switchTotal}_w \le \text{switchMax}$: Barrier packet budget constraint limiting the number of switches per worker. 
\end{itemize}
The execution time $e_k$ is calculated by:
\begin{itemize}
    \item $e_k = \beta_k * (1 + \alpha_k)$: This is the final estimated execution time for a kernel $k$ accounting for slowdown and CU selection. 
    \item $\beta_k = \text{profile}[\sum_c x_{k,c}*c]$: Base solo execution time from profiling. Since $x_{k,c} = 1$ for only one configuration $c$, $\sum_c x_{k,c}*c$ gives the selected CU count for kernel $k$. 
    \item $\alpha_k = \text{CUOverlap}_w/(\text{total \# of CUs})$: Co-location slowdown factor considering CU overlaps with other workers. Previous experiments informed this approach: two identical kernels sharing the same CUs both took twice as long. \footnote{This approach is generalized, but because it is difficult to model we could not resort to a more precise approach.}
\end{itemize}
The CU overlap between workers is tracked by:
\begin{itemize}
    \item $\text{CUAverage}_w = \sum_{k\in w}\sum_c x_{k,c}*c/\text{(\# of kernels in $w$)}$: Average number of CUs per kernel for the worker $w$.
    \item $\text{CUOverlap}_w = \text{CUAverage}_w + \sum_{\forall w':w'\ne w}\text{CUAverage}_{w'}$: Total number of CUs overlapping with other workers.
\end{itemize}

The above formulation is an Integer Linear Programming (ILP) problem which can be solved using ILP solvers like Gurobi. Since the solution time of ILP can be long, we solve this offline for different co-location scenarios and store the results in a lookup table. At runtime, our scheduler simply references this table to get pre-computed CU configurations and assign kernels to the corresponding CU-masked streams.

\begin{figure*}[htbp]
    \centering
    \includegraphics[trim={0 0 0 0}, width=0.5\linewidth, clip]{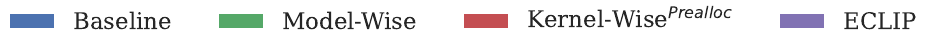}
\end{figure*}

\begin{figure*}[ht!]
    \centering
    \includegraphics[width=1\textwidth]{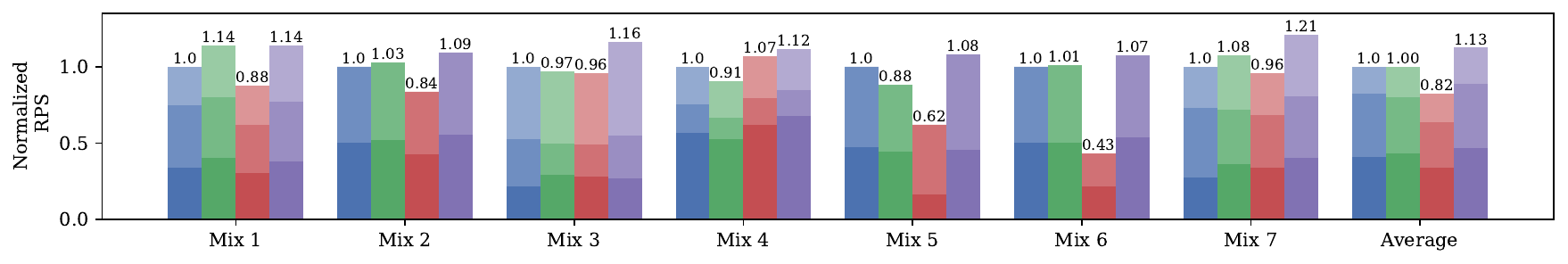}
    \caption[caption]{Normalized Throughput (RPS). Each gradient within the bar represent a different worker.}
    \label{fig:throughputresults}
\end{figure*}



\begin{figure*}[ht!]
    \centering
    \includegraphics[width=1\textwidth]{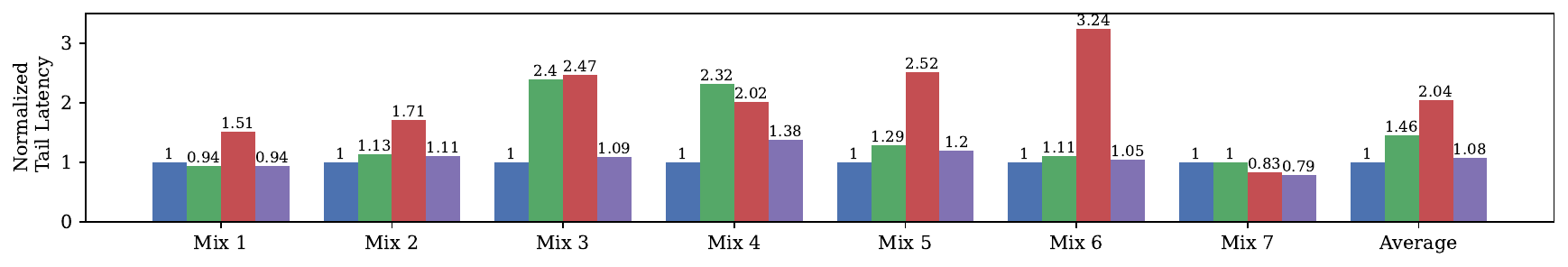}
    \caption{Normalized 95th Percentile Tail Latency}
    \label{fig:latencyresult}
\end{figure*}

\begin{figure*}[ht!]
    \centering
    \includegraphics[width=1\textwidth]{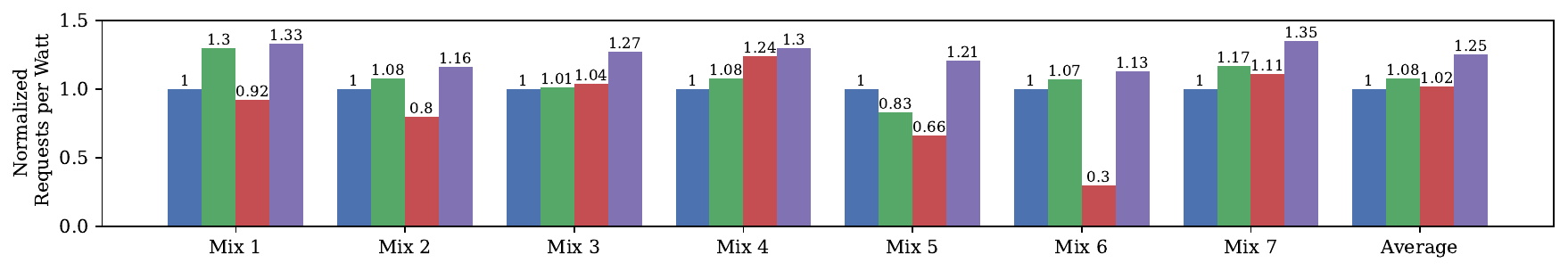}
    \caption{Normalized Energy Efficiency}
    \label{fig:energyefficiency}
\end{figure*}

\section{Evaluation}

\subsection{Experimental Methodology}
Our server consists of two AMD EPYC 7302 16-core CPUs, 512GB memory, and an AMD MI50 GPU. The GPU consists of 60 CUs, arranged as 4 SEs of 15 CUs each. The server runs Ubuntu 20.04 with ROCm 5.2.0 and PyTorch 1.12.0a0. 

We used the \texttt{rocm-smi} library to take power consumption measurements.


\smallskip
\noindent\textbf{Inference Models.}
We evaluate with 7 inference models: \texttt{albert}\cite{albert}, \texttt{densenet201}\cite{densenet201}, \texttt{alexnet}\cite{alexnet}, \texttt{resnet15}\cite{resnet152}, \texttt{resnext101}\cite{resnext101}, \texttt{shufflenet}\cite{shufflenetv2}, and \texttt{vgg19}\cite{vgg19}. 
To stress co-locating ML inference, we test the models at their maximum supported RPS (Requests Per Second). 
We study 7 co-location workload mixes as follows: 
Mix~1 (2 \texttt{albert}, 1 \texttt{densenet201}), 
Mix~2 (1 \texttt{albert}, 1 \texttt{densenet201}), 
Mix~3 (1 \texttt{albert}, 1 \texttt{densenet201}, 1 \texttt{vgg19}), 
Mix~4 (1 \texttt{alexnet}, 1 \texttt{resnet15}, 1 \texttt{vgg19}), 
Mix~5 (1 \texttt{resnext101}, 1 \texttt{shufflenet}),
Mix~6 (2 \texttt{densenet201}),
and Mix~7 (3 \texttt{alexnet}).
We targeted feasible combinations with models of high CU right-sizes~\cite{KRISP} matched with models of low CU right-sizes.

\smallskip\noindent\textbf{Spatial Partitioning Scenarios}
We evaluate four spatial partitioning scenarios as follows: 

\begin{itemize}
\item 
\emph{Baseline}: Unmodified ROCm runtime with no spatial partitioning. Co-located models send all incoming kernels to the default stream that uses all 60 CUs. 

\item 
\emph{Model-Wise}: This approach implements model-wise right-sizing~\cite{gpulet,pariselsa,gslice}. Through profiling, we find each model's minimum number of CUs that satisfies tail latency, and allocate that to each worker for the entire ML inference.

\item 
\emph{Kernel-Wise$^{IOCTL}$} (KW$^{IOCTL}$): This scenario uses CU masking IOCTL to switch CU allocations at \textit{every} kernel, similar to KRISP~\cite{KRISP}. 

\item 
\emph{Kernel-Wise$^{Prealloc}$} (KW$^{Prealloc}$): This scenario switch between pre-allocated CU workers (using ECLIP's mechanisms) for nearly \textit{every} kernel.\footnote{Adjacent kernels often require similar CU counts, not resulting in a switch.} 

\item 
\emph{ECLIP}: Our full ECLIP implementation. 
This configuration accounts for overheads of barrier packets and partitioning, and uses a maximum of 14 partitioning switches per inference request. 14 was selected as it demonstrates a balance between partitioning overhead and co-location efficiency. It is difficult for the model to solve for the number of partitioning switches, so profiling was used to determine this number. \footnote{We experienced diminishing returns with more than around 14 switches, but experienced slowdown with fewer switches.}
\end{itemize}

\begin{table}[!h]
    \centering
    \caption{Sum of per-request CU switches for all workers}
    \label{tab:switchcount}
    \begin{tabular}{|c|c|c|c|c|c|}
    \hline
        Workload & Workers & KW$^{IOCTL}$ & KW$^{Prealloc}$ & ECLIP \\ \hline\hline
        Mix1 & 3 & 1319 & 413 & 42 \\ \hline
        Mix2 & 2 & 1015 & 755 & 23 \\ \hline
        Mix3 & 3 & 1077 & 538 & 28 \\ \hline
        Mix4 & 3 & 613 & 330 & 10 \\ \hline
        Mix5 & 2 & 728 & 166 & 12 \\ \hline
        Mix6 & 2 & 1422 & 695 & 20 \\ \hline
        Mix7 & 3 & 102 & 72 & 30 \\ \hline
    \end{tabular}
\end{table}
As shown in Table~\ref{tab:switchcount}, the number of CU configuration switches done by Kernel-Wise$^{IOCTL}$ is very high, and results in extremely large slowdowns without the custom miroarchitectural support presented in~\cite{KRISP}. Therefore, opted to not graph the results of Kernel-Wise$^{IOCTL}$.



\subsection{Throughput Results}
Figure~\ref{fig:throughputresults} represents the normalized throughput results of the various partitioning scenarios across workload mixes. 
Clearly, across all scenarios, ECLIP achieves the best overall throughput, with a maximum improvement of 21\% for Mix 7 and an average improvement of 13\%. Kernel-Wise$^{Prealloc}$ achieves the lowest throughput due to the overheads of CU switching, as shown in Table~\ref{tab:switchcount}.

\emph{ECLIP}'s fairness modeling attempts to treat each worker with fairness, by not prioritizing the speedup of any one worker. This is best shown with Mix 7 in Figure~\ref{fig:throughputresults}, where the baseline method allows some kernels to cut ahead and finish early despite the workload containing three identical models.

\subsection{Latency Results}
Figure~\ref{fig:latencyresult} shows the normalized 95th percentile tail latency of requests under various spatial partitioning scenarios and workload mixes. 
In general, fine grain partitioning approaches result in higher tail latencies. This is most clearly shown by Kernel-Wise$^{Prealloc}$'s significant increase in tail latency relative to baseline. By limiting the number of switches, our implementation ensures that the tail latency does not rise uncontrollably.
On the other hand, Model-Wise's increase in tail latency occurs from overly coarse grain resource sharing. This is shown best by Mix 3, where the vgg19 worker's kernels typically require all 60 CUs before experiencing slowdown. By forcing the two other workers to constantly contend with vgg19's kernels, all kernels experience slowdown. 

\subsection{Energy Efficiency Results}
Figure~\ref{fig:energyefficiency} shows the normalized energy efficiency. Because of the significant increase in tail latency and switching overheads, Kernel-Wise$^{Prealloc}$'s energy efficiency fails to measurably surpass the baseline. Model-Wise's approach leads to inconsistent improvements due to varying levels of latency and throughput improvements, while ECLIP consistently is able to improve upon the baseline energy efficiency across all mixes by better utilizing the available GPU resources and minimizing switching overheads. ECLIP achieves a maximum energy efficiency improvement of 35\% for Mix 7 with an average of 25\% energy efficiency improvement across all workload mixes. 

\section{Other Related Work}
Previously, we discussed the most relevant work in ML inference servers in Section~\ref{sec:background}. Here, we focus on generic GPU spatial partitioning approaches. Early work on GPU partitioning~\cite{classsearch, warperslicer, improvedscale, saha2019stgm} focused primarily on spatial resource allocation strategies. Later works began to consider partitioning overheads, but with limited depth. \cite{finesharing} acknowledged partitioning overhead without detailed analysis of performance implications. \cite{efficientutil} models partitioning overhead by adding a static slowdown cost in their simulator, but fails to capture the complex behaviors we observed in real systems. While \cite{simulmulti} recognized  hardware context switching overhead, it did not provide detailed analysis of performance impacts or propose solutions to mitigate these overheads. In contrast, our work provides detailed characterization of partitioning overheads on real AMD GPU hardware and presents practical solutions to manage these overheads for ML inference servers.

\section{Conclusion}
This paper demonstrated the drawbacks of current GPU spatial partitioning methodologies and how overly aggressive partitioning leads to severe hardware overhead. Through the use of model-informed compute resource partitioning, our framework, ECLIP, overcomes this issue and significantly improves ML inference throughput and energy efficiency by an average of 13\% and 25\%, respectively, and up to 21\% and 35\%, respectively, while satisfying latency constraints. 


\section*{Acknowledgment}
This work is supported by the National Science Foundation (NSF) grants 1943265, 1955650, 2312395, 2324940, 2324941, and 2047521.



%
\bibliographystyle{IEEEtran}
\bibliography{references}

\end{document}